# RNA Secondary Structure Prediction Using Transformer-Based Deep Learning Models

Yanlin Zhou 1*

Computer Science, Johns Hopkins University,Baltimore,USA
* Corresponding author: popojoyzhou@gmail.com

Tong Zhan 1

Computer Science,Columbia University,NY, USA

tz2483@columbia.edu

Yichao Wu 2

 Computer Science,Northeastern University,Boston, MA,USA

wu.yicha@northeastern.edu

Bo Song 3

 Computer Science,Northeastern University,Boston, MA,USA

song.bo1@northeastern.edu

Chenxi Shi 4

Software development ,Telecommunication Systems Management ,Northeastern University,Boston, MA,USA

shi.che@northeastern.edu

**Abstract**
The Human Genome Project has led to an exponential increase in data related to the sequence, structure, and function of biomolecules. Bioinformatics is an interdisciplinary research field that primarily uses computational methods to analyze large amounts of biological macromolecule data. Its goal is to discover hidden biological patterns and related information. Furthermore, analysing additional relevant information can enhance the study of biological operating mechanisms. This paper discusses the fundamental concepts of RNA, RNA secondary structure, and its prediction.Subsequently, the application of machine learning technologies in predicting the structure of biological macromolecules is explored. This chapter describes the relevant knowledge of algorithms and computational complexity and presents a RNA tertiary structure prediction algorithm based on ResNet. To address the issue of the current scoring function's unsuitability for long RNA, a scoring model based on ResNet is proposed, and a structure prediction algorithm is designed. The chapter concludes by presenting some open and interesting challenges in the field of RNA tertiary structure prediction.

## 1 INTRODUCTION

The computational analysis of RNA sequences is a crucial step in the field of RNA biology. In recent years, machine learning methods have made RNA secondary structure prediction and sequence analysis related to RNA secondary structure more accurate. In addition, AI and machine learning have introduced technological innovations in analyzing RNA-small molecule interactions to discover RNA-targeting drugs and designing RNA aptamers, in which RNA acts as its own ligand. in May 2023, Briefings in Bioinformatics published a review article highlighting recent trends in predicting RNA secondary structure, RNA aptamers, and RNA drug discovery using machine learning, deep learning, and related techniques, and discussing potential future pathways in the field of RNA informatics.

Ribonucleic Acid (RNA) [1]is a carrier of genetic information, which mainly exists in biological cells and some viruses and viroids. RNA performs many complex biological functions in vivo, such as autonomously sensing changes in metabolite concentration, playing a catalytic role and regulating gene expression, and the expression of these functions depends on its tertiary structure. Therefore, the study on the tertiary structure of RNA has become an important research topic. The number of RNA conformations increases exponentially with the increase of the number of nucleotides, and the RNA structures determined by NMR, cryo-electron microscopy and X-ray diffraction are relatively few, with low efficiency and high cost. Therefore, the high-precision RNA tertiary structure prediction algorithm based on biological computation becomes a necessary choice. Currently, popular RNA tertiary structure prediction algorithms include knowledge-based RNA tertiary structure prediction algorithm and physics-based RNA tertiary structure prediction algorithm. Each of these two types of prediction algorithms has its advantages and disadvantages, but neither of them can achieve high-precision and high-integrity RNA tertiary structure modeling. Therefore, the research direction of this paper is to further optimize and improve the relevant prediction algorithm

Deep learning has matured and has achieved great success in many fields such as computer vision [2] and natural language processing [3]. The neural network it uses is a structure with good functionality and versatility, but its effect is highly related to the quantity and quality of the data. However, deep learning methods that have been successful in many fields face challenges in predicting RNA secondary structure. The RNA secondary structure is important for deciphering cell activity and the occurrence of disease. The earliest method used by academics to predict this structure was biological experiments, but this method was too expensive and hindered its popularization. At present, there are no high precision, low data dependence and high convenience RNA secondary structure prediction models. Despite the emergence of efficient and low-cost calculation methods, the accuracy of this calculation method is not satisfactory. Many machine learning methods have also been applied to this area, however, the accuracy rate has not improved significantly. Therefore, more innovative and effective methods are still needed to improve the accuracy and efficiency of RNA secondary structure prediction.

## 2 RELATED WORK

### 2.1 RNA structure prediction

Ribonucleic acid (RNA) carries genetic information in living organisms. Organisms depend on the correct expression of coding RNAs, such as mRNA, and non-coding RNAs, such as [4]tRNA and rRNA, for their routine activities. RNA is involved in all cellular processes and is directly or indirectly linked to disease regulation and occurrence. RNA consists of long chains of molecules, typically four bases connected by phosphodiester bonds. [4]Hydrogen bonds can also form between bases, creating a pair of bases. These pairs can be classified as either normative or non-standard. "Canonical pairing" refers to A base pair in an RNA or DNA molecule where a stable hydrogen bond is formed between adenine (A) and uracil (U) and between guanine (G) and cytosine (C). These are common forms of pairing in biology. The term "irregular pairing" refers to any base pairing other than AU, GC, and GU. These ways of pairing may be less common, but still play a role in RNA or DNA molecules in some cases.

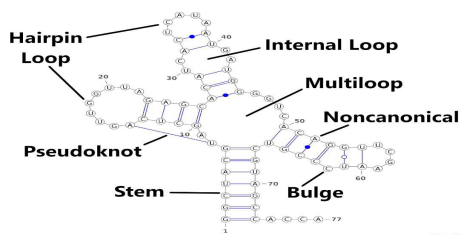

**Figure 1.** Secondary structure diagram of RNA

Figure 1 depicts the secondary structure diagram of RNA, illustrating the intricate folding patterns that result from the convolution and folding of its primary structure. RNA molecules exhibit a quaternary structure, comprising a single strand with base pairs forming the primary structure. The secondary structure consists of various elements such as hairpin loops, stems, internal loops, and pseudoknots, which emerge as a result of the folding process. [5]These secondary structure motifs play crucial roles in RNA function and regulation. Additionally, the tertiary structure of RNA involves further spatial bending of the helix based on the secondary structure, contributing to the molecule's overall three-dimensional conformation. Furthermore, the quaternary structure of RNA involves complex interactions between RNA molecules and proteins, forming intricate complexes essential for various cellular processes.

## 2.2 Deep learning and RNA secondary structure prediction

Recent advancements in deep learning have revolutionized the prediction of RNA and protein structures. For RNA secondary structure prediction, techniques like neural convolutional networks (e.g., ResNet) have shown significant improvements over traditional methods like DCA, doubling the accuracy of internucleotide contact prediction. Similarly, in protein structure prediction, [6]3D convolutional neural networks have gained popularity. These models, like the optimised differentiable model, bridge local and global protein structures by optimizing global geometry while adhering to local covalent chemistry principles, enabling accurate prediction of protein folding structures without prior co-evolution data. In RNA structure prediction, algorithms like [7]FARFAR2 and scoring systems like ARES, based on geometric deep learning, have demonstrated success in blind tests like RNA-Puzzles experiment. Inspired by AlphaFold2's success in protein structure prediction, new approaches like [8]DeepFoldRNA, RoseTTAFoldNA, and RhoFold have emerged. AlphaFold's neural network-based algorithm predicts base pair distances, optimized through gradient descent, enabling the generation of protein structures without complex sampling procedures.

A neural convolutional network is a type of feedforward neural network that uses a convolutional structure based on the mechanism of the visual receptive field. The receptive field refers to a region in the retina of the visual nervous system, and a neuron can only be activated when that region is stimulated. Multiple receptive fields are interlaced and overlapping, eventually covering the entire line of sight.

Convolutional layers, also known as convolution layers, and fully connected layers are the basic structural units of convolutional neural networks. They possess the structural characteristics of pooling, shared weights, and local receptive fields. In comparison to fully connected networks, convolutional neural networks are capable of performing operations such as spatial translation and rotation. This not only preserves the internal correlation of the data but also reduces the relevant parameters in the network model. The convolutional structure effectively reduces the probability of overfitting the model.

The ResNet residual unit can be expressed as:

$$y_l = h(x_l) + F(x_l, W_l) \quad x_{l+1} = f(y_l) \quad h(x_l) = x_l \quad (10$$

Where l represents the L-th residual unit, and xl and xl+1 represent its input and output, respectively. F() represents the residual function, f( ) represents the Relu activation function, and there are many types of Relu functions.

The main focus of algorithm analysis is to assess its correctness and complexity. [9]Correctness is the fundamental criterion for evaluating an algorithm, which is achieved when the algorithm produces the correct output after a series of limited and clear instructions when given an example problem. Complexity analysis is another crucial factor in evaluating algorithm performance. Algorithm complexity analysis typically involves evaluating the space-time complexity of algorithms. Time complexity refers to the total number of times the algorithm's basic operation is executed during its execution, while space complexity refers to the amount of memory required during implementation.

## 2.3 Transformer automatically predicts RNA structure

The emergence of the Transformer neural network, driven by an attention mechanism, has revolutionized structural biology. In 2020, DeepMind's AlphaFold2 marked a significant breakthrough by accurately predicting the three-dimensional structure of proteins from their amino acid sequences. This framework, powered by a Transformer neural network, excels in capturing long-range dependencies within input sequences, going beyond their sequential neighborhood. Subsequent frameworks such as [10]RoseTTAFold and OmegaFold have further advanced protein structure prediction, building upon the success of AlphaFold2.

To objectively evaluate emerging protein structure prediction methods, the prediction community acknowledges the importance of blind tests. The recently concluded CASP15 conference provided a valuable platform for assessing these approaches. In a study published in PNAS, researchers benchmarked the predictive modeling performance of 69 CASP15 single-chain protein targets. They conducted fully automated modeling using open-source software implementations of various methods and evaluated the accuracy of full-length predictions compared directly to experimental coordinates, without segmenting them into domains. Additionally, domain-level analyses were performed on multidomain proteins to assess the accuracy of individual domains and their structures, utilizing static databases, libraries, and model weights published before CASP15.

The Transformer model employs probabilistic autoencoders and [11]ELBO optimization to maximize marginal likelihood. Initially, the input sequence is encoded to generate vectors representing it. These vectors undergo processing using the Probabilistic Transformer model to produce a probability distribution representing the likelihood of each target tag given the input. During inference, the prediction model can yield varying results depending on the sample taken. In practice, the Transformer operates as an Encoder-Decoder architecture, with the middle portion divided into encoding and decoding components.

The Transformer architecture revolutionized natural language processing (NLP) tasks by introducing a self-attention mechanism, enabling it to capture long-range dependencies in sequences efficiently. In practice, the Transformer architecture can indeed be viewed as an Encoder-Decoder architecture, consisting of two main components: the encoder and the decoder.

1. Encoder Component: The encoder is responsible for processing the input sequence, such as a sentence or a document, and transforming it into a sequence of contextualized representations. Each token in the input sequence is passed through multiple layers of self-attention mechanisms and feed-forward neural networks, allowing the model to capture the contextual information of each token in relation to the entire input sequence. The output of the encoder is a sequence of feature vectors, each representing a token in the input sequence.

2. Decoder Component[12]: The decoder takes the output from the encoder and generates the output sequence, typically in a token-by-token fashion. It also consists of multiple layers, each equipped with self-attention mechanisms and feed-forward neural networks. However, in addition to attending to the input sequence like the encoder, the decoder also attends to the encoder's output through a mechanism called encoder-decoder attention. This allows the decoder to focus on relevant parts of the input sequence while generating the output tokens. At each step, the decoder predicts the next token in the output sequence based on the previously generated tokens and the encoder's output.

By dividing the Transformer architecture into these two components, the model can effectively handle various sequence-to-sequence tasks, such as machine translation, text summarization, and question answering. The encoder learns to encode the input sequence into a fixed-length representation, capturing its semantic meaning, while the decoder uses this representation to generate the output sequence. This modular design has proven to be highly effective and flexible for a wide range of NLP tasks.

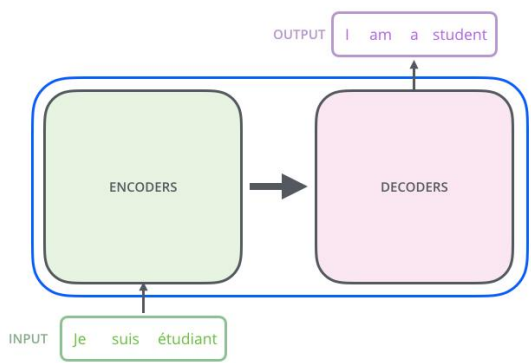

Figure 2.Transformer model (Encoder-Decoder architecture pattern)

The encoding component of the Transformer architecture consists of a multi-layer Encoder, which in this paper employs six layers. Each encoder layer contains two sub-layers(figure 2): a Self-Attention layer and a Position-wise Feed Forward Network (FFN)[13]. In the Self-Attention layer, the encoder can leverage information from other words in the input sentence to encode a specific word, allowing it to focus not only on the current word but also on contextual information from surrounding words. The output from the Self-Attention layer is then passed to the feedforward network for further processing.

Similarly, the decoding component of the Transformer architecture also consists of decoders with six layers, each containing Self-Attention and FFN sub-layers. Additionally, there is an Attention layer (encoder-decoder Attention) between these sub-layers, enabling the decoder to focus on relevant parts of the input sentence, akin to the attention mechanism in seq2seq models.

The Transformer architecture's suitability for addressing challenges in predicting RNA structures stems from two key features. Firstly, it can accurately model long-term dependencies in sequence data by incorporating positional encoding into the input sequence. This allows the model to capture remote dependencies between input features without interference from intervening features. Secondly, the Transformer architecture is adept at modeling unordered sets of entities and their interactions, which is challenging for many other deep learning architectures. This is achieved by conducting most operations in a positional manner, enabling the model to handle unordered sets of features effectively. These advantages make the Transformer architecture an attractive choice for quantitative modeling of histone codes, as it enables researchers to simultaneously consider multiple remote regulatory regions near the transcription start site (TSS) in the wider genomic window.

## 3   EXPERIMENT AND METHODOLOGY

trRosettaRNA, a tool for predicting RNA structure, has two main steps. First, it uses a technique called transformer networks to predict the one - and two-dimensional geometries of RNA. It then minimizes the energy to convert these geometries into the three-dimensional structure of RNA. By benchmarking its performance, trRosettaRNA was found to work better than traditional automated methods. In some blind tests, trRosettaRNA's predictions matched the top predictions of human experts. As a result, trRosettaRNA also outperforms other deep learning-based methods in many structural prediction competitions. Still, there are challenges in using automated methods to predict the precise structure of synthetic RNA. The researchers hope that this work will provide a good starting point for solving the problem of predicting RNA structure.

### 3.1   Experimental data set

This paper proposes the use of a Transformer network as an automatic sequence prediction model to predict RNA nucleotide sequences. To systematically generate different nucleic acid secondary structures, traditional neural network rMSA and SPOT-RNA program are employed. The generated secondary structure is then converted into a model, and the MSA representation and pair representation are modified. The resulting initial transformer network, called RNAformer, is used to predict 1D and 2D geometry. At the core of these steps is the ability to translate the geometry of the generated structural model into constraints to guide the final step in the folding of the gene structure based on energy minimization.

### 3.2 Data processing procedure

The complete process of transformer networks in RNA secondary structure prediction involves three main steps. The first is the input data preparation phase. At this stage, for a given query RNA sequence, a large MSA (multiple sequence alignment) is first generated from multiple sequence databases (such as NCBI's nt, Rfam, and RNAcentral) by using rMSA programs. Second, the final MSA is selected by running the Infernal program against the smaller RNAcentral database and based on the quality of the predicted distance graph. At the same time, SPOT-RNA was used to predict the secondary structure of RNA.

The next step is to predict one - and two-dimensional geometries. In this step, a transformer network called RNAformer is used, similar to the Evoformer network in AlphaFold2. The RNAformer network first converts the input MSA and secondary structure into two representations: the MSA representation and the pair representation. Each RNAformer block then updates these two representations through four steps: (1) MSA to MSA, (2) MSA pairing, (3) pairs of two, and (4) pairing to MSA. In single-channel RNAformer, 48 blocks are looped 4 times in a complete inference, and finally the two-dimensional geometric prediction probability distribution is obtained by linear layer and softmax operation.

The final step is to generate the all-atomic structure model. In this step, based on the predicted one - and two-dimensional geometry information, the corresponding algorithms and models are used to generate a model of the all-atomic structure of RNA. This process may involve parsing, optimizing, and validating the predicted results to obtain a final high-quality structural model.

Similar to trRosetta, trRosettaRNA uses deep learning potentials and physics-based energy terms in Rosetta to generate a complete model of atomic structure by minimizing the energy defined below:

$$E = \omega_1 E_{dist} + \omega_2 E_{ori} + \omega_3 E_{cont} + \omega_4 E_{ros} \quad (1)$$

$$E_{ori} = E_{ori,2D} + \frac{L}{2} E_{ori,1D} \quad (2)$$

The folding process in pyRosetta involves generating 20 all-atomic starting structures for each RNA using RNA_HelixAssembler, followed by refinement through quasi-Newtonian optimization L-BFGS to minimize total energy. The total energy comprises constraints based on distance (Edist), direction (Eori), contact (Econt), and Rosetta's internal energy term. Constraints in 2D and 1D directions are represented by Eori,2D and Eori,1D, respectively. The weights (w1 = 1.03, w2 = 1.0, w3 = 1.05, w4 = 0.05) are determined to minimize the average RMSD, based on hundreds of randomly selected RNAs from the training set. After refinement, 20 finely refined all-atomic structure models are obtained for each RNA, from which the model with the lowest total energy (Eq.1) is chosen as the final prediction.

### 3.3 Experimental results explain

1.trRosettaRNA's performance on 30 individual RNAs trRosettaRNA was tested against two other methods, RNAComposer and SimRNA, using 30 RNA structures. The average deviation in structure prediction (RMSD) was 8.5 angstroms for trRosettaRNA, compared to 17.4 angstroms for RNAComposer and 17.1 angstroms for SimRNA. This significant difference highlights trRosettaRNA's superior performance in RNA structure prediction.

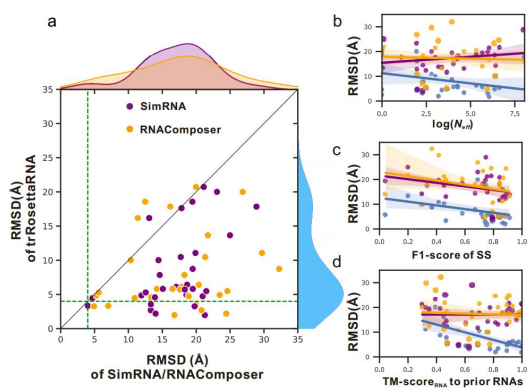

**Figure 2.** Results of RNA2D structure prediction model

The results from the RNA2D structure prediction model, as depicted in Figure 2, highlight the superior performance of trRosettaRNA compared to traditional methods such as RNAComposer and SimRNA. In a dataset comprising 30 instances, trRosettaRNA outperformed RNAComposer 86.7% of the time and SimRNA 96.7% of the time. Notably, 20% of the models generated by trRosettaRNA exhibited an RMSD (Root Mean Square Deviation) of less than 4 A, a level of accuracy that neither RNAComposer nor SimRNA could achieve. [14]These findings underscore the efficacy of trRosettaRNA in RNA structure prediction, surpassing the capabilities of conventional methods.

The Das group was identified as the most accurate method, submitting models for 17 targets. It's worth noting that while some participating groups may leverage human expertise or literature data for guidance, trRosettaRNA's predictions are entirely automated and demonstrate comparable accuracy to the top-performing human prediction group. Overall, these results highlight the effectiveness of trRosettaRNA in addressing the challenges of RNA structure prediction and its potential to rival expert human predictions.

2. Blind test of CASP15

In the blind testing of CASP15, the researchers participated as part of the Yang-Server team, utilizing the trRosettaRNA model as an automation server. They achieved a significant 9th position among 42 RNA structure prediction teams, including both human and server teams. Notably, within the server teams, Yang-Server ranked second, following UltraFold_Server. Furthermore, Yang-Server's performance was enhanced when considering the cumulative Z-score (> 0.0) of RMSD, placing 5th among all groups and 1st among server groups. Remarkably, Yang-Server surpassed other deep learning-based groups in terms of Z-scores for RMSD. Particularly accurate predictions were observed for two protein-binding targets, R1189 and R1190, highlighting the method's potential in predicting protein-binding RNAs, despite the absence of binding partner information, though further accuracy improvements are possible.

Furthermore, blind tests were conducted on recent RNA-Puzzles targets, where the authors participated as part of the automated server group called Yang. These targets included PZ37 (Ligand-bound dimer), PZ38 (Ligand-bound ribose switch), and PZ39 (A protein-binding clover RNA). For PZ37 and PZ38, the results were highly competitive, with the author's models ranking 3rd out of 16 and 15 participating groups, respectively. However, for target PZ39, the model's performance was less satisfactory, with an RMSD greater than 15 A. This can be attributed to the unique characteristics of PZ39, which lacks sequence and structural similarity to known RNAs, posing a challenge for the method's performance on this particular objective.

To address these challenges and improve future RNA structure predictions, one potential approach is to integrate the method with traditional techniques and optimize algorithms for underrepresented RNA structures. For instance, utilizing neural networks, such as physics-based neural networks, to learn force fields or identify/assemble local patterns instead of directly predicting global[14] 3D structures could mitigate biases against known RNA folding and enhance prediction accuracy.

## 4 CONCLUSION

Machine learning techniques, particularly deep learning approaches, have shown promising results in predicting RNA secondary structures. Compared to thermodynamic model-based methods, deep learning approaches make fewer assumptions, allowing for the consideration of false knots, third-order interactions, non-standard base pairing, and other previously unidentified constraints. Experimental findings indicate that the prediction accuracy of the multi-objective evolutionary strategy algorithm surpasses that of the evolutionary strategy algorithm, especially for long RNA sequences. However, the prediction performance for some complex RNA secondary structures remains suboptimal. This may be attributed to the inaccuracy of the selected external interface for computing free energy, which affects the accurate calculation of free energy for complex RNA secondary structures. Additionally, in some cases, the real structure of complex RNA secondary structures may be predominantly determined by corresponding points in the target space, resulting in predicted structures that are extremely close to reality but not accurately predicted.

Future improvements in the prediction performance of the algorithm can be achieved by refining the mutation operator and fitness evaluation function based on the multi-objective evolutionary strategy algorithm. Furthermore, exploring alternative free energy computing interfaces and refining the false knot free energy computation method could enhance the prediction accuracy of the algorithm. These advancements are crucial for addressing the challenges posed by complex RNA secondary structures and improving the overall efficacy of machine learning-based RNA structure prediction methods.

## REFERENCES

[1] Turner, Douglas H., Naoki Sugimoto, and Susan M. Freier. "RNA structure prediction." Annual review of biophysics and biophysical chemistry 17.1 (1988): 167–192.

[2] Seetin, Matthew G., and David H. Mathews. "RNA structure prediction: an overview of methods." Bacterial regulatory RNA: methods and protocols (2012): 99–122.

[3] Reuter, Jessica S., and David H. Mathews. "RNAstructure: software for RNA secondary structure prediction and analysis." BMC bioinformatics 11 (2010): 1–9.

[4] Wang, Yong, et al. "Construction and application of artificial intelligence crowdsourcing map based on multi–track GPS data." arXiv preprint arXiv:2402.15796 (2024).

[5] Zhou, Y., Tan, K., Shen, X., & He, Z. (2024). A Protein Structure Prediction Approach Leveraging Transformer and CNN Integration. arXiv preprint arXiv:2402.19095.

[6] Zhang, Chenwei, et al. "Enhanced User Interaction in Operating Systems through Machine Learning Language Models." arXiv preprint arXiv:2403.00806 (2024)

[7] Ni, Chunhe, et al. "Enhancing Cloud–Based Large Language Model Processing with Elasticsearch

and Transformer Models." arXiv preprint arXiv:2403.00807 (2024).

[8] Shapiro, Bruce A., et al. "Bridging the gap in RNA structure prediction." Current opinion in structural biology 17.2 (2007): 157–165.

[9] Gardner, Paul P., and Robert Giegerich. "A comprehensive comparison of comparative RNA structure prediction approaches." BMC bioinformatics 5 (2004): 1–18.

[10] Miao, Zhichao, and Eric Westhof. "RNA structure: advances and assessment of 3D structure prediction." Annual review of biophysics 46 (2017): 483–503.

[11] Zheng, Jiajian, et al. "The Random Forest Model for Analyzing and Forecasting the US Stock Market in the Context of Smart Finance." arXiv preprint arXiv:2402.17194 (2024).

[12] Yang, Le, et al. "AI-Driven Anonymization: Protecting Personal Data Privacy While Leveraging Machine Learning." arXiv preprint arXiv:2402.17191 (2024).

[13] Cheng, Qishuo, et al. "Optimizing Portfolio Management and Risk Assessment in Digital Assets Using Deep Learning for Predictive Analysis." arXiv preprint arXiv:2402.15994 (2024).

[14] Wu, Jiang, et al. "Data Pipeline Training: Integrating AutoML to Optimize the Data Flow of Machine Learning Models." arXiv preprint arXiv:2402.12916 (2024).